\PassOptionsToPackage{table}{xcolor}

\documentclass[pdflatex,sn-nature]{sn-jnl}

\usepackage{graphicx}%
\usepackage{multirow}%
\usepackage{amsmath,amssymb,amsfonts}%
\usepackage{amsthm}%
\usepackage{mathrsfs}%
\usepackage[title]{appendix}%
\usepackage{textcomp}%
\usepackage{manyfoot}%
\usepackage{booktabs}%
\usepackage{algorithm}%
\usepackage{algorithmicx}%
\usepackage{algpseudocode}%
\usepackage{listings}%

\usepackage[table]{xcolor}%
\usepackage{colortbl}%

\definecolor{B1}{RGB}{222,235,247}
\definecolor{B2}{RGB}{158,202,225}
\definecolor{B3}{RGB}{107,174,214}
\definecolor{B4}{RGB}{49,130,189}
\definecolor{B5}{RGB}{8,81,156}
\definecolor{R1}{RGB}{254,224,210}
\definecolor{R2}{RGB}{252,174,145}
\definecolor{R3}{RGB}{251,106,74}
\definecolor{R4}{RGB}{222,45,38}
\definecolor{R5}{RGB}{165,15,21}
\newcommand{\negcell}[2]{%
  \ifdim #2pt < 0.10pt \cellcolor{B1}%
  \else\ifdim #2pt < 0.20pt \cellcolor{B2}%
  \else\ifdim #2pt < 0.30pt \cellcolor{B3}%
  \else\ifdim #2pt < 0.40pt \cellcolor{B4}%
  \else \cellcolor{B5}\color{white}%
  \fi\fi\fi\fi #1}
\newcommand{\poscell}[2]{%
  \ifdim #2pt < 0.10pt \cellcolor{R1}%
  \else\ifdim #2pt < 0.20pt \cellcolor{R2}%
  \else\ifdim #2pt < 0.30pt \cellcolor{R3}%
  \else\ifdim #2pt < 0.40pt \cellcolor{R4}%
  \else \cellcolor{R5}\color{white}%
  \fi\fi\fi\fi #1}
\newcommand{\fdrcell}[1]{\fbox{#1}}

\begin{document}

\title[EEG Entrainment to Non-Speech Rhythmic Patterns Associated with Dimensions of Language Performance]{EEG Entrainment to Non-Speech Rhythmic Patterns Associated with Dimensions of Language Performance}


\author[1,3,4]{\fnm{Francisco J.} \sur{Martínez-Murcia}}\email{fjesusmartinez@ugr.es}
 
\author*[2,3]{\fnm{Nicolás J.} \sur{Gallego-Molina}}\email{ngjm@uma.es}
\equalcont{These authors contributed equally to this work.}

 \author[1,3]{\fnm{Marco A.} \sur{Formoso}}\email{marco.a.formoso@ugr.es}
\equalcont{These authors contributed equally to this work.}

\author[2,3]{\fnm{Andrés} \sur{Ortiz}}\email{andres.ortiz@uma.es}
\equalcont{These authors contributed equally to this work.}
 
\affil*[1]{\orgdiv{Department of Signal Theory, Networking and Communications}, \orgname{University of Granada}, \orgaddress{ \city{Granada}, \postcode{18071}, \country{Spain}}}
 
\affil[2]{\orgdiv{Department of Communications Engineering}, \orgname{University of Málaga}, \orgaddress{ \city{Málaga}, \postcode{29004}, \country{Spain}}}
 
\affil[3]{\orgdiv{Andalusian Research Institute in Data Science and Computational Intelligence (DaSCI)}, \orgaddress{ \country{Spain}}}

\affil[4]{\orgdiv{Research Institute in Information and Communications Technology (CITIC-UGR)} , \orgaddress{ \city{Granada},\country{Spain}}}


\abstract{
Developmental dyslexia (DLX) affects 5--12\% of the global population and impairs learning to read. Its biological basis, particularly the difficulties observed in oral phonological tasks, remains debated. Entrainment, the coupling of brain oscillations to speech rhythms, is essential for encoding phonological representations. We investigated whether brain entrainment to non-speech rhythms could predict DLX in 48 participants listening to white noise amplitude-modulated at rates matching Spanish speech units. Phase-amplitude coupling (PAC) analysis of electroencephalography (EEG) signals yielded accurate DLX prediction for the 40~Hz stimulus using the Heights Ratio PAC measure (balanced accuracy 0.929, $p = 0.0069$, in the best case), indicating a stronger phonemic entrainment deficit in dyslexic individuals. Regression analysis in control (CTL) participants showed a strong dependence between reading performance scores and greater entrainment in language-associated brain regions. In particular, participants with poorer phonemic awareness, reading comprehension, or rapid automatized naming showed greater right-hemisphere entrainment to the 40~Hz stimulus. This finding adds evidence for the hemispheric compensatory mechanisms hypothesized in the literature. In conclusion, our methodology identifies dyslexia-specific entrainment patterns near language-associated brain regions, together with compensatory entrainment in right-hemisphere areas.
}

\keywords{Phase Amplitude Coupling, Entrainment, EEG, Learning Disorders}



\maketitle

\section{Introduction}\label{sec:introduction}

Reading and writing are key skills in today's society. Poor reading and writing performance has a clear effect on self-esteem and school failure \cite{terras2009dyslexia}, and has therefore been the subject of study for a long time. When low performance in these abilities is not associated with low intelligence, poor vision or inadequate educational opportunities, we speak of dyslexia (DLX). The real prevalence of DLX is unknown, but current estimates speak of between 5 and 12$\%$ of the global population \cite{Peterson2012}, across different languages, cultures, and socioeconomic backgrounds. DLX affects different aspects of reading skills \cite{ziegler2005reading}: a difficulty in recognizing and manipulating the sound structure of words (phonological awareness); challenging the connection of letters to their corresponding sounds (decoding); a slow and laborious reading, affecting comprehension and enjoyment of written material (fluency); persistent difficulties in accurately spelling words (spelling) and struggling with retaining and processing information while reading (working memory). Therefore, it poses significant challenges in educational systems worldwide, as traditional teaching methods may not adequately address the unique learning needs of dyslexic individuals.

The neurobiological underpinnings of these phonological deficits can be better understood within the framework of current theory of language \cite{hickok2007cortical,friederici2013language}. According to these theories, the encoding of audio information into words is achieved at a neural level by a distributed network oscillating at specific temporal rates. This process is commonly known as ``entrainment'' \cite{molinaro2016out}. Entrainment to speech rhythms has been proven using techniques such as magnetoencephalography (MEG) or electroencephalography (EEG) in subjects under a wide variety of stimuli \cite{kuhl2004early,hickok2007cortical,giraud2013neurogenetics,friederici2013language,molinaro2016out,kimppaImpairedNeuralMechanism2018}.  Abnormal entrainment between neural oscillations and speech rates can also result in an inability to segment speech correctly into hierarchical units such as stress, syllables, or phonemes. This conforms to the core phonological deficits found in DLX. However, there is evidence that phonological awareness (PA) can be trained \cite{Temple2003}. Consequently, a more objective and earlier diagnosis of DLX may pave the way to reducing its psychosocial impact. Differences in entrainment to speech have been assessed by estimating the coupling between brain rhythms (in EEG or MEG) and specific speech units \cite{kimppaImpairedNeuralMechanism2018,goswamiSpeechRhythmLanguage2019,goswamiNeuralOscillationsPerspective2019,virtala2020poor,Serniclaes2021}. Specifically, studies have reported impaired synchrony of natural speech to the delta band (0.5-1 Hz) in adults \cite{molinaro2016out}, impaired low-frequency entrainment to natural speech in children \cite{DiLiberto2018}, and reduced entrainment to natural and vocoded speech in the delta band (0-2 Hz) \cite{POWER20161}. In \cite{goswamiSpeechRhythmLanguage2019}, 
Goswami et al. revealed that children with DLX might rather show atypical synchronization of cortical oscillations in the theta range (4-8Hz). Two recent studies have showed atypical delta band oscillations in dyslexic children exposed to syllabic stimuli and natural speech \cite{keshavarzi2022atypical,keshavarzi2022decoding}. Abnormal patterns were also found in the low gamma (25-35 Hz) bands in \cite{LEHONGRE20111080}. 

These works prove that entrainment to speech rhythm is a potential marker for the detection of DLX. Closely linked to the entrainment phenomenon, a powerful measure of stimulus-driven neural modulation is the phase-amplitude cross-frequency coupling (PAC), which has been used successfully with both children and adults \cite{molinaro2016out,Flanagan2018,DiLiberto2018,LEHONGRE20111080}. PAC assesses whether the phase of low-frequency oscillations modulates the amplitude envelope of high frequency oscillations. In the brain, PAC has been shown to measure effective communication during cognitive processes \cite{lakatos2005oscillatory}, revealing a hierarchical structure of the EEG data. It has also been used to model the connectivity of brain regions in which theta oscillation triggers high gamma power \cite{canolty2006high}. Measures such as heights ratio (HR) \cite{lakatos2005oscillatory}, the Kullback-Leibler modulation index (KL-MI) \cite{tort2010measuring}, mean vector length (MVL) \cite{canolty2006high}, gaussian copula PAC (gcPAC) \cite{ince2017statistical} or phase-locking value (PLV) \cite{hulsemann2019quantification} have been proposed. In fact, the phase synchrony used in some studies with children and adults with DLX can be considered a simplification of PLV \cite{goswamiSpeechRhythmLanguage2019,Flanagan2018,virtala2020poor}. 

With a growing corpus of machine learning (ML) research over the last decade, there have also been many papers applying statistical learning approaches to the diagnosis of developmental dyslexia. The very first approach we know of used EEG-derived features and a linear classifier, namely a support vector machine classifier (SVC) as in  \cite{andreadis2009dislexiaApEn}. Other works computed the Event Related Potential (ERP) of the EEG channels, and classified the features also using SVC \cite{frid2012dislexiaSVM}. Perera et al. tested SVC on EEG signals for the same task, but also linear discriminant analysis (LDA) and Neural Networks \cite{perera2016review}, concluding favorably to SVC. Most of these methods use raw speech signals or AM carriers -an audible tone modulated with the amplitude envelope of real speech- as stimuli. However, there is increasing evidence that the impaired PA may arise as a bottom-up cascade effect caused by impaired low-level audio processing \cite{giraud2012cortical,virtala2020poor,ortiz2020dyslexia}. In this context, non-speech audio stimuli at typical speech rates may allow the identification of impaired auditory processing leading to impaired PA. Some works attempted to classify EEG obtained from children with dyslexia using temporal and spectral features of subjects under non-speech auditory stimuli \cite{ortiz2020dyslexia,formoso2021detecting,martinezmurciaEEGConnectivityIJNS2020}, showing that there are relevant differences between children with dyslexia and normal readers, and even relating some of these associations to indicators of PA. 

In this work, we have extensively explored this possibility. We synthesized white-noise amplitude-modulated at different rates related to speech units of the Spanish language: a stimulus at 4.8 Hz for syllables, 16 Hz for intrasyllabic segmentation rhythms (e.g., the /p/ and /a/ in /pa/, but also the /pl/ and /\ae{}n/ in /pl\ae{}n/), and 40 Hz for phonemes. Quantifying the stimulus-driven PAC of the brain to these synthetic signals may allow us to investigate whether there is a non-linguistic, impaired entrainment to acoustic rhythms. This approach differs from classical entrainment measures, which quantify phase-locking of EEG oscillations to a periodic stimulus; the proposed stimulus-driven PAC approach instead measures how much a EEG band amplitude is organised by the low-frequency stimulus phase, yielding a cross-frequency coupling index that reflects stimulus-driven modulation of neural activity. We conducted the study in 48 7-year-old children, 15 of whom have been diagnosed with DLX, and recorded EEG signals while listening to the stimulus. We then measured the entrainment between the stimulus and the EEG using five different PAC measures: HR, KL-MI, MVL, gcPAC and PLV.
This stimulus-driven coupling stems from the instantaneous phase of the synchronized stimulus (\(\phi_{1}(t)\)) and the instantaneous amplitude of the narrow-band EEG signal (\(a_{2}(t)\)) at different electrodes, both extracted using the Hilbert transform (HT). Each narrow-band EEG signal is produced by applying a FIR bandpass filter at 49 different frequency points, spanning from the stimulus frequency up to 200 Hz, distributed on a logarithmic scale. In order to discard the possibility that a high PAC was produced by chance we employed permutation testing by generating surrogate versions of the amplitude signal by barrel-switching and computing the surrogate PAC measures in each case. From this surrogate PAC distribution, Z-scores were derived, providing an estimate of the statistical significance of the observed PAC measure, \(PAC_o\).

For each PAC measure, the resulting PAC Z-scores were later aggregated per EEG band and used to generate features for a SVC. The aggregation per EEG frequency band is performed by creating a set of PAC Z-scores for each band, $PAC_{band}$, in which the PAC Z-scores in the range of the frequency extrema of the band are included only if they are above the significant Z-score threshold 1.65. The final PAC for each subject is the average of the normalized $PAC_{band}$ for every 5-second segment. Then, the diagnostic accuracy of these PAC-band features is assessed using a linear SVC in a repeated stratified 5-fold cross-validation scheme.

\begin{figure}[!h]
\centering
\includegraphics[width=0.9\textwidth]{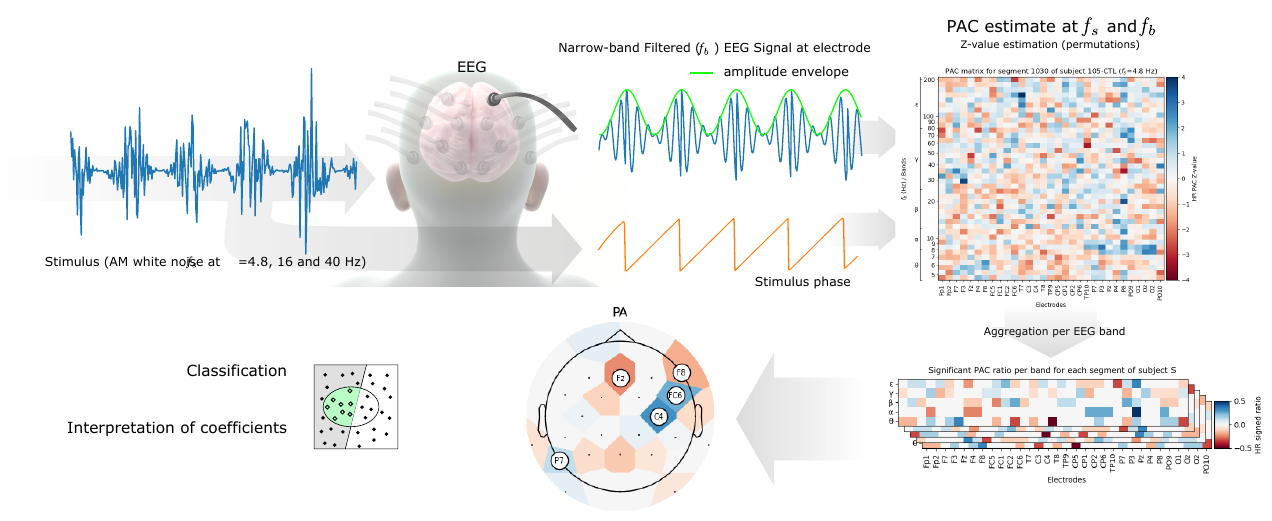}
\caption{Experimental setup for the acquisition, preprocessing and computation of PAC features used throughout this work, and how they are aggregated and pooled to inform the explainable classification system.}\label{fig:experimental-setup}
\end{figure}

In a previous stage, the children's reading performance was assessed by means of several questionnaires focused on reading skills, oral skills and phonological awareness. Considering this information, we explored the relationships between the aggregated PAC features and these aspects of reading performance and analyzed their specific impacts using a Multi-task Lasso model, which applied a mixed L1/L2 regularization norm. Finally, the experimental setup is shown in Figure \ref{fig:experimental-setup} and the main contributions of this work are: the proposal of non-speech entrainment for automatic DLX detection; the study of a wide variety of PAC measures in this task; the permutation-based aggregation feature to obtain the most significant CFC features and the exploration and interpretation of the interrelation between the PAC features that yield high performance in a machine learning system and the dimensions of reading performance.

\section{Results}\label{sec:results}

\subsection{Heights Ratio is an Optimal CFC Measure for Dyslexia}

\begin{figure}[h]
\centering
\includegraphics[width=1\textwidth]{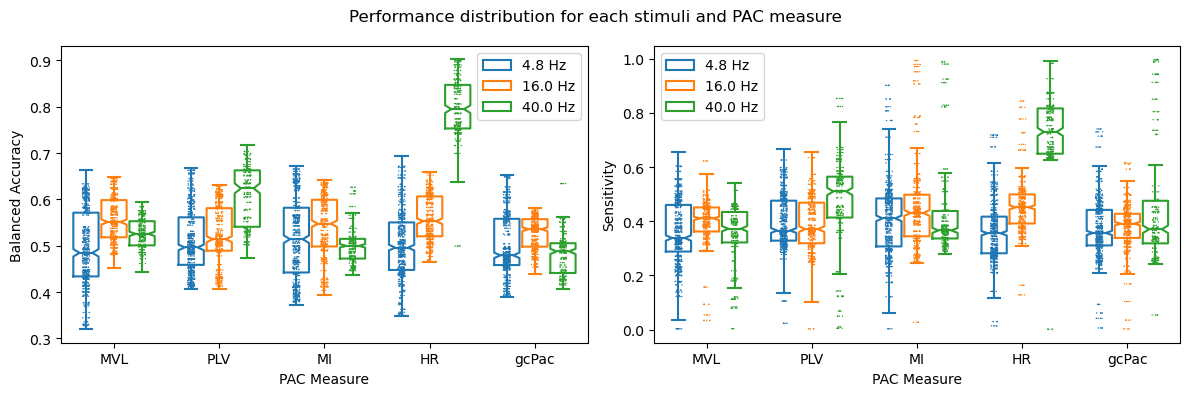}
\caption{Comparison of performance measures in experiments for each stimuli and PAC estimation method. Each point represents the cross-validated balanced-accuracy / sensitivity for a SVC with a different combination of EEG band, parameter C and SVC penalty using all 31 EEG channels.}\label{fig:swarmplot}
\end{figure}

Our classification analysis reveals that HR is by far the best PAC measure for the stimuli and segments, achieving up to 90$\%$ balanced accuracy for the HR and 40 Hz stimulus (mean accuracy and [standard deviation] over all CV folds in the best case: 0.929 [0.08], $p=0.0069$). Figure \ref{fig:swarmplot} shows the distribution of the performance achieved by each PAC measure, as estimated by balanced accuracy (trade off between sensitivity and specificity) and sensitivity. Each point represents a performance estimate with a different combination of EEG band, parameter C of SVC and SVC penalty using all 31 EEG channels. This allows the estimation of the best performing measures and stimuli in discriminating subjects with and without DLX.

While the 40 Hz stimulus is the best performing one, when measured with the HR, the 16 Hz stimulus also achieves high performance with HR (around 80$\%$, best balanced acc. 0.830 [0.11], $p=0.0225$), and slightly less with KL-MI (around 70$\%$ sensitivity and balanced accuracy). As for the 4.8 Hz stimulus, all PAC measures achieve similar -and worse- sensitivity, but specificity is slightly higher in the case of PLV and MVL. The best accuracy in this case is not significant (0.738 [0.13], $p=0.0945$).

\subsection{Entrainment to 40 Hz stimulus Accurately Predicts Dyslexia} \label{sec:class_weight}

\begin{figure}[h]
\centering
\includegraphics[width=0.6\textwidth]{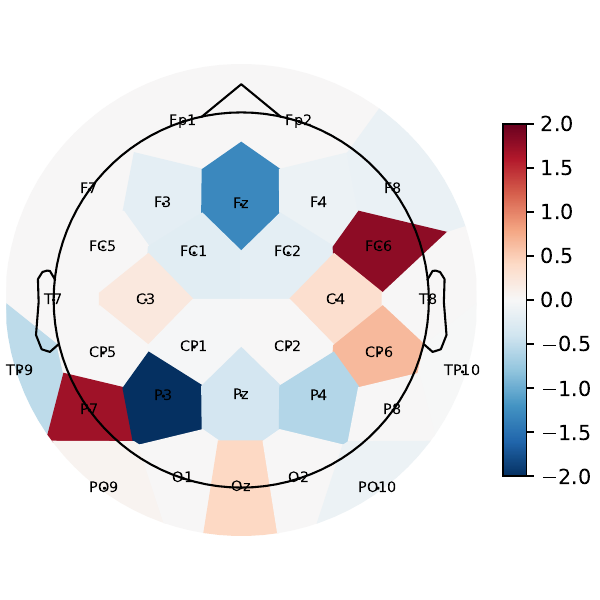}
\caption{Classification weights of the linear SVC trained for discriminating between DLX
and CTL subjects using the HR entrainment between the 40 Hz stimulus and Gamma band.}\label{fig:coef_topo}
\end{figure}

Regarding the combination of stimulus and band, the entrainment between the 40 Hz stimulus and the gamma band, with HR is the most discriminant. To dig deeper into this performance, we depict the SVC coefficients -linear weights vector- for these two stimuli and the different bands, measured by HR at Figure \ref{fig:coef_topo}.

This figure illustrates that the contribution of the electrodes in the two hemispheres is not symmetric for 40 Hz gamma-band entrainment. The largest positive coefficients are found at FC6 and P7. In particular, in the right hemisphere there is a line formed by FC6-C4-CP6 that contributes positively to the classification as DLX. There are large negative coefficients at electrodes located around key language areas in the left hemisphere: P3, TP9, FC1 and F3. In addition, it is observed that some of their respective electrodes in the right hemisphere P4, F4 and FC2 and the central line with Fz and Pz also have a negative contribution. This revealed that high entrainment in large areas of the right hemisphere (FC6, C4, CP6) contributed for a subject to be classified as DLX. On the other hand, larger entrainment at specific regions of the left hemisphere (F3, FC1, P3) contributed for a subject to be considered in the CTL group.

Therefore, these classification weights expose differential patterns of PAC entrainment to the stimulus that have a greater contribution in the right hemisphere to classification as DLX for the most discriminative entrainment revealed by the linear SVC.

\subsection{Heights Ratio Correlates with Dimensions of Reading Performance}

The reading performance of the study children was evaluated using a series of questionnaires addressing reading, oral, and phonological performance. Of these, we explore whether the aggregated HR-PAC features relate to these dimensions of reading performance and what kind of influence they make. 

\begin{figure}[h]
\centering
\includegraphics[width=1\textwidth]{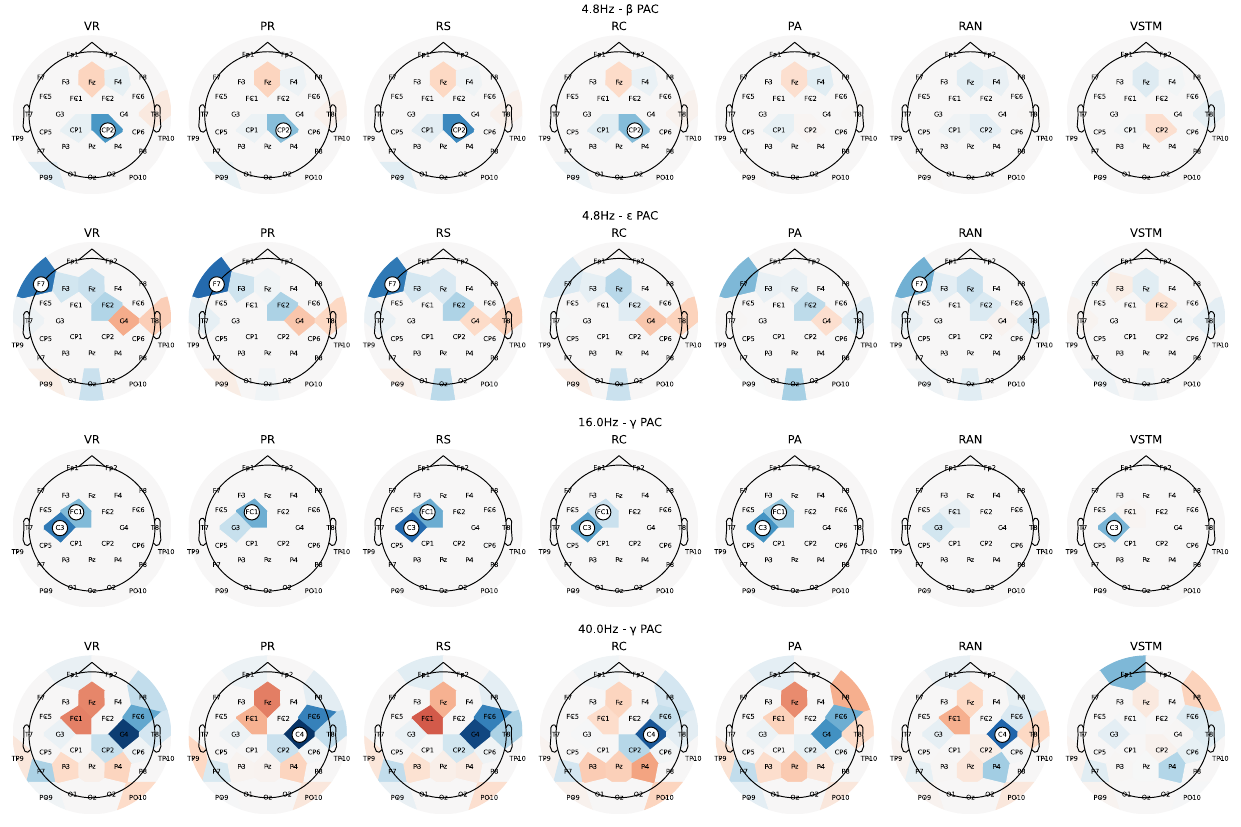}
\caption{Location of estimated coefficients for the relationship between the stimulus and the frequency band PAC at different channels and the measured behavioral variables.  Only electrodes with significant multiple-comparison-corrected $p$-values are highlighted. Columns correspond to the seven behavioural indices (VR, PR, RS, RC, PA, RAN, VSTM); rows correspond to stimulus--band combinations.}\label{fig:topoplots}
\end{figure}

These relationships have been computed using a Multi-task Lasso model trained with the L1/L2 mixed-norm as regularizer, fitted by coordinate descent \cite{hastie2015statistical}. The empirical statistical significance of the coefficients was calculated by permutation testing. We generated a null-distribution of the coefficient matrix by permuting the target variables (reading performance test outcomes) 99999 times and estimated the $p$-values for the two-sided hypothesis. To address multiple comparisons across 31 channels and 7 behavioural indices, we applied Benjamini--Hochberg False Discovery Rate (FDR-BH) corrections. The estimated coefficients and significant channels that survived the correction are highlighted at Figure \ref{fig:topoplots}.

Two of the PAC coefficient maps displayed at Figure \ref{fig:topoplots} are particularly of interest, given the diversity and number of significant relationships: the 4.8Hz - $\epsilon$ and especially, the 40Hz - $\gamma$ entrainment. The coefficients for this fast-rhythm entrainment are in agreement with the areas exhibited by the classification weights of the linear SVC trained to discriminate between DLX and CTL subjects using the HR measure of the 40 Hz - $\gamma$ entrainment in Section \ref{sec:class_weight}. For this case, the most relevant coefficients values and their corrected p-value are shown in Table \ref{tab:correlation}. The variables related to reading performance (VR, PR, RS and RC) have many points in common. In the low-rhythm entrainment to the $\epsilon$ band, there is strong negative correlation with F7, located over the left frontal region of the scalp, i.e, roughly around Broca's area. This region is associated with language processing, working memory, and aspects of executive function. 

The negative correlation observed may be associated with an altered network for phonological processes reflecting a dysfunctional cortical mechanism which is in line with the neuroimaging literature \cite{spironelliDysfunctionalHemisphericAsymmetry2008,shaywitz2003neural,shaywitzFunctionalDisruptionOrganization1998}. These altered mechanisms are also observed for the 40Hz - $\gamma$ entrainment in which most of the subjects show a strong negative correlation between the reading performance and C4 (significant after FDR correction), FC6, and generally, all pre-motor cortex and nearby areas (e.g., the right hemisphere regions that corresponds to Broca's area in the left hemisphere). This is consistent with the idea of a compensatory mechanism for dyslexic reading located in the right hemisphere \cite{shaywitz2003neural} to compensate for the failure of the posterior brain systems of the left hemisphere to function properly \cite{shaywitzDisruptionPosteriorBrain2002}. Thus, we found that higher values of the 40Hz - $\gamma$ entrainment measured with PAC in these areas of the right hemisphere are associated with lower scores on reading performance variables.

On the other hand, a positive strong correlation was found in FC1 and Fz for the 40Hz - $\gamma$ entrainment, approximately placed around left pre-motor cortex. These results support the left dominant entrainment to acoustic modulations at frequencies associated with the phonemic sampling frequency found in  controls \cite{LEHONGRE20111080} since higher PAC values in these EEG channels correspond to higher reading performance.

In relation to this, patterns associated with PA can also be found in both 4.8Hz - $\epsilon$ and the 40Hz - $\gamma$ entrainment. For slow-rhythms, a strong negative relationship between the PAC values measured and the PA score has been found. This may indicate an altered activation of the phonological processing networks measured with PAC reflecting the atypical processing in dyslexia suggested in the TSF \cite{goswamiTemporalSamplingFramework2011}. Additionally, the 40Hz - $\gamma$ entrainment exhibits a link between PA and entrainment in the right hemisphere, consistent with the aforementioned compensatory mechanism.

\begin{table*}[ht]
\centering
\caption{Coefficients of the Multi-task Lasso model for the aggregated PAC gamma
entrainment with the 40~Hz stimulus between EEG electrodes and behavioural indices.
Cell colour matches the 40~Hz--$\gamma$ topoplot convention:
\textcolor{B4}{\textbf{blue}} = negative coefficient;
\textcolor{R4}{\textbf{red}} = positive coefficient.
Electrodes surviving FDR-BH correction are additionally marked in bold and their corrected p-values (in brackets) are also shown, in which significant results $(p<0.05)$ and $(p<0.01)$ are denoted with * and ** respectively.}
\label{tab:correlation}
\setlength{\fboxsep}{1pt}
\resizebox{\textwidth}{!}{%
\begin{tabular}{@{}lcccccccc@{}}
\toprule
\textbf{Electrode} & \textbf{VR} & \textbf{PR} & \textbf{RS} & \textbf{RC} & \textbf{PA} & \textbf{RAN} & \textbf{VSTM} \\
\midrule
Fp1
  & \negcell{$-$0.036}{0.036}
  & \negcell{$-$0.029}{0.029}
  & \negcell{$-$0.056}{0.056}
  & \negcell{$-$0.016}{0.016}
  & \negcell{$-$0.050}{0.050}
  & \negcell{$-$0.037}{0.037}
  & \negcell{$-$0.223}{0.223} \\
Fz
  & \poscell{$+$0.243}{0.243}
  & \poscell{$+$0.257}{0.257}
  & \poscell{$+$0.178}{0.178}
  & \poscell{$+$0.113}{0.113}
  & \poscell{$+$0.235}{0.235}
  & \poscell{$+$0.103}{0.103}
  & \poscell{$+$0.051}{0.051} \\
F8
  & \negcell{$-$0.122}{0.122}
  & \negcell{$-$0.053}{0.053}
  & \negcell{$-$0.048}{0.048}
  & \negcell{$-$0.093}{0.093}
  & \poscell{$+$0.181}{0.181}
  & \negcell{$-$0.042}{0.042}
  & \poscell{$+$0.114}{0.114} \\
FC1
  & \poscell{$+$0.253}{0.253}
  & \poscell{$+$0.177}{0.177}
  & \poscell{$+$0.305}{0.305}
  & \poscell{$+$0.090}{0.090}
  & \poscell{$+$0.115}{0.115}
  & \poscell{$+$0.180}{0.180}
  & \negcell{$-$0.007}{0.007} \\
FC6
  & \negcell{$-$0.258}{0.258}
  & \negcell{$-$0.337}{0.337}
  & \negcell{$-$0.342}{0.342}
  & \negcell{$-$0.120}{0.120}
  & \negcell{$-$0.243}{0.243}
  & \negcell{$-$0.076}{0.076}
  & \negcell{$-$0.069}{0.069} \\
\textbf{C4}
  & \negcell{$-$0.474 [0.097]}{0.474}
  & \negcell{\fdrcell{$-$0.489 [0.034]*}}{0.489}
  & \negcell{$-$0.456 [0.097]}{0.456}
  & \negcell{\fdrcell{$-$0.428 [0.040]*}}{0.428}
  & \negcell{$-$0.301 [0.250]}{0.301}
  & \negcell{\fdrcell{$-$0.402 [0.016]*}}{0.402}
  & \poscell{$+$0.005 [1.000]}{0.005} \\
P7
  & \negcell{$-$0.158}{0.158}
  & \negcell{$-$0.081}{0.081}
  & \negcell{$-$0.124}{0.124}
  & \negcell{$-$0.093}{0.093}
  & \negcell{$-$0.123}{0.123}
  & \negcell{$-$0.042}{0.042}
  & \negcell{$-$0.057}{0.057} \\
P4
  & \poscell{$+$0.112}{0.112}
  & \poscell{$+$0.113}{0.113}
  & \poscell{$+$0.040}{0.040}
  & \poscell{$+$0.199}{0.199}
  & \poscell{$+$0.060}{0.060}
  & \negcell{$-$0.174}{0.174}
  & \negcell{$-$0.147}{0.147} \\
\bottomrule
\end{tabular}}
\end{table*}

\section{Discussion}

Our results provide robust evidence of a link between neural oscillations and abnormal reading in children. We explored the entrainment of the brain to three different stimuli as measured by five different PAC strategies, aggregated the entrainment per EEG frequency band, and used the resulting features as input to a classifier distinguishing DLX from CTL subjects. This approach consistently achieved accuracy above 90\% for the entrainment between a 40~Hz stimulus and the gamma band, as measured by HR (mean balanced accuracy 0.929, $p = 0.0069$). To our knowledge, this represents the highest classification performance reported in EEG-based dyslexia detection studies to date, surpassing previous approaches that achieved accuracies ranging from 70\% to 85\% using temporal, spectral, or connectivity features \cite{andreadis2009dislexiaApEn, frid2012dislexiaSVM, ortiz2020dyslexia, formoso2021detecting, martinezmurciaEEGConnectivityIJNS2020}.

A central finding of this work is the superior discriminative power of the Heights Ratio compared to other PAC measures. The entrainment between the 16~Hz stimulus and gamma band measured with HR also achieved accuracy above 80\% (0.830, $p = 0.0225$), despite the classifier hyperparameters chosen. Therefore, HR is the best performing measure, followed by KL-MI. We propose that this difference stems from the computational properties of each measure. Both HR and KL-MI depend on the quantization of phases of the instantaneous phase vector, as defined in Eq.~\ref{eq:vquant}. HR is particularly better at detecting large differences in the quantized amplitude vector, as can be seen from Eq.~\ref{eq:hr}. Larger differences imply a larger modulation index, which in our case is better captured by the simpler HR measure. In contrast, KL-MI estimates divergence from uniformity, which may be less sensitive to extreme modulation patterns that characterize the dyslexia-related entrainment deficit.

On the other hand, MVL has poorer performance. As it was demonstrated to be a good choice for high-quality segments \cite{canolty2010functional}, it may be that its performance in real-world noisy environments like ours is diminished. A similar conclusion can be extracted from PLV. Note that PLV and MVL work better with the 4.8~Hz stimulus. This may be due to several reasons, but possibly due to a higher resolution of the phase stimulus (more than 100 points per cycle, versus about 12 points per cycle for the 40~Hz stimulus at $F_s = 500$~Hz). Higher resolution in the phase allows for a higher-resolution of the PAC estimation, and that benefits MVL and PLV, although their performance is still limited ($< 0.7$ accuracy). For its part, gcPAC assumes gaussianity on the signals. However, this assumption does not appear to hold after closer examination of the acquired EEG data, which may explain its suboptimal performance.

The 40~Hz stimulus, corresponding to the average phonemic rate in Spanish speech \cite{Lizarazu2015}, emerged as the most discriminative stimulus for identifying dyslexia. This finding aligns with and extends the Temporal Sampling Framework (TSF) proposed by Goswami \cite{goswamiTemporalSamplingFramework2011}, which posits that dyslexia arises from atypical neural sampling of the speech amplitude envelope at multiple temporal rates. According to the TSF, neural oscillations at delta ($\sim$2~Hz), theta ($\sim$5~Hz), and low-gamma ($\sim$30~Hz) frequencies are responsible for parsing speech into stress patterns, syllables, and phonemes, respectively. Our results are consistent with this framework insofar as we observed the strongest group differences at the phonemic rate. Giraud and Poeppel \cite{giraud2012cortical} proposed that gamma-band oscillations ($\sim$30--50~Hz) are essential for encoding phonemic-level information during speech processing, and that fast-rate information in lower gamma should be key for detecting distinctive features in predictive speech processing. As a consequence, an increased sensitivity towards phonemes in dyslexics would explain the difficulty to establish this encoding into language-level phoneme categories \cite{kuhl2004early, giraud2013neurogenetics}. Furthermore, an impaired gamma entrainment could lead to difficulties in segmenting phonemes and mapping them onto linguistic categories, which are fundamental processes for reading acquisition \cite{LEHONGRE20111080}.

Crucially, because this deficit was detected using non-linguistic stimuli (amplitude-modulated white noise), our findings suggest that the core impairment involves basic auditory temporal processing mechanisms rather than speech-specific representations. This interpretation has important implications for early identification and intervention. If the entrainment deficit is fundamentally auditory, it may be detectable before the onset of reading instruction, potentially enabling earlier intervention. Moreover, auditory training paradigms targeting temporal processing might prove beneficial as a complement to phonological interventions. Additionally, the 40~Hz stimulus is closely related to the 40~Hz steady-state auditory response (ASSR). When the stimulation rate is 40~Hz, constructive interference occurs, causing the resulting signal to have a higher amplitude than the original signal, which can facilitate the detection of neural response objectively. This constructive interference is due to the fact that when the stimulus is presented with an inter-stimulus interval of 25~ms (40~Hz rate), the Na-Pa components of one response overlap with the Nb-Pb components of the adjacent response, resulting in a higher amplitude signal \cite{bohorquezGeneration40HzAuditory2008}. This physiological amplification may facilitate the detection of between-group differences in neural synchronization, partially explaining why the 40~Hz stimulus outperformed both the 4.8~Hz (syllabic) and 16~Hz (intrasyllabic) conditions.

The 16~Hz stimulus also achieved notable classification performance (83.0\% accuracy), suggesting that intrasyllabic segmentation processes may also be affected in dyslexia. This intermediate rate corresponds to the temporal structure of consonant-vowel transitions and onset-rime boundaries, which are critical for grapheme-phoneme mapping \cite{goswamiSpeechRhythmLanguage2019}. These results indicate that multiple levels of the temporal hierarchy may be compromised in dyslexia, with phonemic-rate processing showing the most pronounced deficit. These differential patterns of PAC entrainment to the stimuli found by the linear SVC trained for discriminating between DLX and CTL subjects are related to reading performance scores, as evidenced by the estimated coefficients of the Multi-task Lasso model. A striking finding from the SVC weight analysis is the asymmetric contribution of left and right hemisphere electrodes to dyslexia classification. The figure illustrating classification weights shows that the contribution of the electrodes in the two hemispheres is not symmetrical for 40~Hz gamma-band entrainment. The largest positive coefficients are found at FC6 and P7. In particular, in the right hemisphere there is a line formed by FC6-C4-CP6 that contributes positively to the classification as DLX. There are large negative coefficients at electrodes located around key language areas in the left hemisphere: P3, TP9, FC1 and F3. In addition, some of their respective electrodes in the right hemisphere (P4, F4, FC2) and the central line with Fz and Pz also have a negative contribution. This revealed that high entrainment in large areas of the right hemisphere (FC6, C4, CP6) contributed for a subject to be classified as DLX, whereas larger entrainment at specific regions of the left hemisphere (F3, FC1, P3) contributed for a subject to be considered in the CTL group. 

This hemispheric pattern admits several interpretations that warrant careful consideration. The negative correlation between right-hemisphere PAC values and reading performance scores is consistent with the hypothesis that right-hemisphere regions may be recruited to compensate for left-hemisphere dysfunction in dyslexia \cite{shaywitz2003neural, hoeft2011neural}. According to this view, individuals with dyslexia engage right-hemisphere homologues of language areas (e.g., right inferior frontal regions corresponding to Broca's area) as an adaptive response to left-hemisphere deficits. The right hemisphere often compensates for deficits in the left hemisphere across a wide range of unilateral language impairments, providing an adaptive mechanism for maintaining function \cite{kellHowBrainRepairs2009, preibischEvidenceCompensationStuttering2003}. However, this form of neural reorganization rarely leads to complete behavioral recovery. In the context of dyslexia, such compensation may further disrupt the typical lateralization of phonemic processing, potentially exacerbating difficulties in phonological encoding and impairing language development. This compensatory recruitment appears to be functionally inefficient, as indicated by its association with poorer reading outcomes.

An alternative explanation is that the observed pattern reflects atypical lateralization of auditory processing rather than active compensation. Studies have reported reduced leftward asymmetry or even rightward asymmetry in language-related brain regions in individuals with dyslexia \cite{spironelliDysfunctionalHemisphericAsymmetry2008}. Spironelli et al. reported inverted anterior hemispheric lateralization in dyslexia, which suggests functional impairments in language processing. Under this view, the right-hemisphere entrainment may represent a primary deficit in establishing typical left-hemisphere dominance for phonological processing, rather than a secondary compensatory response. It is important to note that SVC weights reflect the contribution of features to a classification boundary, not direct measures of neural activity or its functional significance. The observed patterns should therefore be interpreted as markers that distinguish groups rather than as direct evidence of underlying neural mechanisms.

In general, individuals with dyslexia consistently struggle with tasks involving phonological awareness, rapid lexical access, and verbal short-term memory. Evidence suggests that this difficulty stems from a core phonological deficit linked to the quality of phonological representations or their accessibility and retrieval \cite{ramusTheoriesDevelopmentalDyslexia2003} that hinders the performance of phonological awareness tasks and the learning of sound-letter mapping. Supporting this hypothesis, findings from 40~Hz stimulus--gamma entrainment reveal a pattern of right-hemisphere negative correlation and left/central positive correlation with reading performance scores. This aligns with research indicating that language performance in dyslexia is partially compensated by neural mechanisms in the right hemisphere, as observed in both adults \cite{shaywitz2003neural} and adolescents \cite{hoeft2011neural}. Thus, we found that higher values of the 40~Hz--gamma entrainment measured with PAC in these areas of the right hemisphere are associated with lower scores on reading performance variables.

On the other hand, a positive strong correlation was found in FC1 and Fz for the 40~Hz--gamma entrainment, approximately placed around left pre-motor cortex. These results support the left dominant entrainment to acoustic modulations at frequencies associated with the phonemic sampling frequency found in controls \cite{LEHONGRE20111080}, since higher PAC values in these EEG channels correspond to higher reading performance. The variables related to reading performance (VR, PR, RS and RC) have many points in common. In the low-rhythm entrainment to the epsilon band, there is strong negative correlation with F7, located over the left frontal region of the scalp, roughly around Broca's area. This region is associated with language processing, working memory, and aspects of executive function. The negative correlation observed may be associated with an altered network for phonological processes reflecting a dysfunctional cortical mechanism, which is in line with the neuroimaging literature \cite{spironelliDysfunctionalHemisphericAsymmetry2008, shaywitz2003neural, shaywitzFunctionalDisruptionOrganization1998}. This finding may point to increased activation in regions crucial for grapheme-phoneme conversion, potentially reflecting compensatory mechanisms that serve as markers of impaired reading in individuals with dyslexia. In relation to this, patterns associated with PA can also be found in both 4.8~Hz--epsilon and the 40~Hz--gamma entrainment. For slow-rhythms, a negative relationship between the PAC values measured and the PA score has been found. This may indicate an altered activation of the phonological processing networks measured with PAC, reflecting the atypical processing in dyslexia suggested in the TSF \cite{goswamiTemporalSamplingFramework2011}. Additionally, the 40~Hz--gamma entrainment exhibits a link between PA and entrainment in the right hemisphere, consistent with the aforementioned compensatory mechanism.

In both children and adults with typical reading abilities, auditory statistical learning has been found to have a stronger connection to reading proficiency compared to visual statistical learning \cite{qiHearingMattersMore2019}. In children, this relationship is further influenced by phonological awareness, an essential emergent literacy skill. Implicit learning of auditory sequences may therefore serve as a foundational step toward developing phonological awareness, which plays a crucial role in literacy acquisition. Ozernov-Palchik et al. \cite{ozernov-palchikIntactProceduralMemory2023} found that the association between auditory statistical learning and reading-related abilities was more pronounced in dyslexic adults compared to typical readers. This indicates that the relationship between auditory statistical learning and reading may be especially significant in situations that require greater decoding effort, such as in individuals with dyslexia, young children, and tasks involving nonword reading. The specificity of reduced auditory statistical learning in individuals with dyslexia suggests an underlying deficit in low-level learning mechanisms, which could affect both auditory learning and reading proficiency. Atypical auditory learning may contribute to difficulties in establishing grapheme-phoneme associations, a fundamental skill for reading. Neurophysiological evidence supports this hypothesis: reading difficulties may stem from inconsistencies in the auditory system's response to sound, which in turn alters the ability to link phonology and orthography during reading. Hornickel and Kraus \cite{hornickelUnstableRepresentationSound2013} indicated that poor readers show significantly greater variability in auditory brainstem responses to speech compared to proficient readers, regardless of baseline neurophysiological noise levels. This greater neural variability may be a fundamental biological factor contributing to the well-documented behavioral and neural deficits observed in individuals with reading difficulties.

Differences in phoneme gamma entrainment may therefore serve as a potential biomarker for detecting dyslexia. The fact that dyslexia-related entrainment deficits can be detected using amplitude-modulated white noise suggests that the underlying impairment involves basic auditory processing mechanisms rather than speech-specific representations. This supports the hypothesis that phonological deficits in dyslexia may arise as a downstream consequence of impaired low-level auditory temporal processing \cite{giraud2013neurogenetics, virtala2020poor}.

Compared to MEG-based studies \cite{molinaro2016out, DiLiberto2018}, our EEG approach offers a practical advantage for clinical translation due to lower cost and greater accessibility. While MEG provides superior spatial resolution, our results demonstrate that meaningful group differences can be detected with standard EEG equipment, supporting the feasibility of developing scalable diagnostic tools. The entrainment deficits revealed in this work are consistent with the altered segmentation processes during language acquisition \cite{noordenbosNeuralEvidenceAllophonic2012} and coherent with the well-known behavioral deficits of children with dyslexia \cite{ziegler2005reading}. Furthermore, these findings are supported by the state-of-the-art performance achieved in classification, being best in terms of performance among those using EEG, a cheaper alternative than MRI and MEG.

Despite the limited spatial resolution of low-density EEG, the observed patterns such as left hemisphere correlations for slow rhythms and premotor region correlations for the 40~Hz stimulus, align with areas known to be involved in reading and language processing. This analysis corroborates the utility and interpretability of the proposed methodology, since the PAC measures cannot only be used for differential diagnosis, but also to provide understanding on the neurobiological basis of phonological processes related to the reading performance scores via the Multi-task Lasso model.

Several limitations of this study should be acknowledged. The sample of 48 children (15 with dyslexia) is modest, though comparable to similar studies in this field \cite{molinaro2016out, DiLiberto2018, power2013neural}. Recruitment of subjects for research involving children with dyslexia is often challenging, and this study faced similar obstacles. The class imbalance (31$\%$ dyslexic) reflects the population prevalence but may affect classifier training. We addressed this through stratified cross-validation and balanced accuracy metrics. To verify that this sample provides adequate statistical power for the primary classification analysis, we conducted a post-hoc power analysis using the observed effect size. The most discriminative electrode in the HR–40 Hz–gamma feature space was FC6 (Cohen's d = 1.62), followed by P3 (d = 1.27) and CP6 (d = 1.22), all corresponding to very large effect sizes by conventional benchmarks. Using the effect size at FC6, a two-group power analysis accounting for the unequal group sizes in our sample (ratio = 33/15) and setting $\alpha$ = 0.05 and 1-$\beta$ = 0.80 requires approximately 5 participants in the smaller group, a threshold comfortably exceeded by our 15 dyslexic participants. Furthermore, the use of repeated stratified 5-fold cross-validation (50 repetitions) and permutation-based significance testing mitigates the risk of inflated performance estimates that can arise in small samples, with the observed permutation p-value of 0.0069 confirming that results are unlikely to reflect chance. Furthermore, the electrodes exhibiting the largest effect sizes (FC6, P3 and CP6) are neurophysiologically coherent with the SVC classification weights shown in Figure \ref{fig:coef_topo}, which independently identified those same electrodes as the most discriminative. Thus, this convergence between the effect size ranking and the classifier weights provides an additional layer of internal validation for the robustness of the findings despite the modest sample size.

The study focused solely on Spanish speakers. Spanish has a relatively transparent orthography compared to languages like English, which may influence both the manifestation of dyslexia and the stimulus rates most relevant for detection. Thus, the generalizability of the findings to other languages remains to be explored. All participants were 7-year-old children at a specific developmental stage. The neural entrainment patterns associated with dyslexia may evolve with age and reading experience. Longitudinal studies and cross-sectional comparisons across age groups would clarify developmental trajectories. EEG has limited spatial resolution, and the electrode positions provide only approximate localization of neural sources. The hemispheric patterns we observed are suggestive but should not be interpreted as precise anatomical localization. Source reconstruction techniques or multimodal imaging would strengthen anatomical inferences. Our correlational design cannot establish whether atypical entrainment is a cause, consequence, or correlate of reading difficulties. Longitudinal studies beginning before literacy instruction, combined with intervention studies, would help clarify causal relationships. Finally, to address multiple comparisons in the Multi-task Lasso analysis (31 channels $\times$ 7 indices), we applied FDR-BH correction. Nevertheless, given the exploratory nature of the analysis and the modest sample size, associations that survive FDR correction should still be interpreted with caution, and replication in independent samples is needed.

\section{Conclusions}

This study demonstrates that neural entrainment to non-speech auditory rhythms, quantified through stimulus-driven PAC, can accurately distinguish children with dyslexia from typical readers. We investigated whether brain entrainment to non-speech rhythms could serve as a biomarker for developmental dyslexia by analyzing PAC between amplitude-modulated white noise stimuli at rates matching Spanish speech units and EEG signals in 48 children.

Among the five PAC measures evaluated (HR, KL-MI, MVL, gcPAC, and PLV), the HR consistently outperformed alternatives across all stimulus conditions. The combination of HR with the 40~Hz stimulus and gamma-band EEG activity achieved 92.9\% balanced accuracy ($p = 0.0069$), representing the highest classification performance reported to date among EEG-based dyslexia detection studies employing entrainment or PAC features with repeated stratified cross-validation. This finding establishes HR as the preferred PAC measure for this application and suggests that dyslexia-related entrainment deficits manifest as pronounced asymmetries in phase-amplitude distributions. The superior performance of HR likely stems from its sensitivity to extreme modulation patterns, which appear to characterize the entrainment deficit in dyslexia.

The 40~Hz stimulus, corresponding to the average phonemic rate in Spanish, yielded superior classification performance compared to syllabic (4.8~Hz) and intrasyllabic (16~Hz) rates. This result provides direct support for the TSF hypothesis that phoneme-level temporal processing is particularly compromised in dyslexia. The 16~Hz stimulus also achieved notable performance (83.0\% balanced accuracy), suggesting that multiple levels of the temporal hierarchy may be affected, with phonemic-rate processing showing the most pronounced deficit. Crucially, because these deficits were detected using non-linguistic stimuli, our findings suggest that the core impairment involves basic auditory temporal processing mechanisms rather than speech-specific representations, a distinction with important implications for early detection and intervention strategies.

Classification weights and regression coefficients revealed a consistent hemispheric pattern: elevated right-hemisphere gamma entrainment (electrodes FC6, C4, CP6) was associated with dyslexia classification and poorer reading performance, whereas elevated left-hemisphere and midline entrainment (electrodes F3, FC1, Fz, P3) was associated with typical reading. This asymmetry is consistent with theories of compensatory right-hemisphere recruitment in dyslexia, though alternative interpretations involving primary lateralization deficits cannot be excluded. Regardless of the underlying mechanism, this hemispheric signature offers a neurophysiologically interpretable marker for reading difficulties.

The Multi-task Lasso analysis demonstrated that entrainment patterns are not uniformly related to reading ability but show specificity across performance dimensions. The 40~Hz--gamma entrainment at electrode C4 was the strongest predictor across reading variables (visual route, phonological route, reading speed, reading comprehension) and phonological awareness. The 4.8~Hz--epsilon entrainment at F7 (approximating Broca's area) specifically correlated with reading performance measures. These differential associations suggest that PAC features capture distinct aspects of the neural circuitry supporting reading, enhancing both the interpretability and potential clinical utility of the approach.

Beyond the empirical findings, this work introduces methodological innovations that may benefit future research. The use of amplitude-modulated white noise at speech-relevant rates provides a language-neutral probe of auditory temporal processing, potentially enabling cross-linguistic comparisons and earlier (pre-literacy) assessment. Our approach of retaining only statistically significant PAC values before band aggregation reduces noise and enhances the reliability of features entering the classifier. The comprehensive evaluation of five PAC measures under identical conditions provides empirical guidance for measure selection in future entrainment studies. The combination of linear SVC for classification and Multi-task Lasso for regression yields models whose coefficients can be mapped onto electrode locations and behavioral dimensions, supporting neurophysiological interpretation.

The high classification accuracy achieved with a relatively simple paradigm (passive listening to modulated noise) and accessible technology (standard EEG) suggests potential for developing objective, scalable screening tools. Unlike behavioral assessments, which require verbal responses and may be confounded by attention, motivation, or test anxiety, neural entrainment measures could provide a complementary, more direct index of auditory processing capacity. The ability to use non-speech stimuli is particularly valuable for pre-literacy assessment and for populations where language-based testing is challenging. If phonemic entrainment deficits reflect a fundamental auditory processing limitation, then interventions targeting temporal processing (such as rhythmic training, musical training, or auditory-based approaches) may offer a useful complement to traditional phonological interventions. Furthermore, PAC measures could potentially serve as biomarkers for tracking intervention response, providing an objective index of neural change that may precede or accompany behavioral improvement.

Future research should prioritize validation in larger, independent, and linguistically diverse cohorts to establish the generalizability and clinical utility of these findings. Longitudinal studies tracking entrainment from pre-literacy through reading acquisition would clarify developmental trajectories and the predictive value of early entrainment measures. Intervention studies testing whether entrainment patterns can be modified and whether such changes predict reading outcomes would address questions of causality and therapeutic potential. Multimodal studies combining EEG with neuroimaging techniques offering complementary spatial resolution would help clarify the anatomical substrates of the observed effects.

In conclusion, this study establishes that PAC between non-speech auditory rhythms and neural oscillations (particularly at the phonemic rate and measured with the HR) can accurately identify children with developmental dyslexia and predict specific dimensions of reading performance. The methodology is grounded in current theories of speech processing and dyslexia, yields interpretable neurophysiological markers, and utilizes accessible EEG technology. While validation in larger and more diverse samples is required, these findings represent a meaningful step toward objective, early identification of reading difficulties, which remains essential for timely intervention and improved outcomes for the millions of children affected by dyslexia worldwide.

\section{Methods}\label{sec:methods}

\subsection{Experimental Design}
The dataset used in this work was obtained by the LEEDUCA research group at the University of Málaga (Spain) \cite{ortiz2020dyslexia}. The LEEDUCA cohort follows more than 1400 children aged 4 to 8 years, applying quarterly a complete battery of cognitive and linguistic tasks whose content is based on recent studies \cite{de2017longitudinal}. The study was approved by the Medical Ethical Committee of the University of Malaga (05/02/2020 PND016/2020) and supported by the education office of the Junta de Andalucía (regional government) which granted permission to evaluate students at different public schools. 

These assessments evaluated, among others, reading, oral and phonological performance. Reading performance is evaluated in four different categories: reading of one, two and three-syllable words -visual route (VR)-, reading of one, two and three- syllable pseudowords -phonological route (PR)-, number of words per minute -reading speed (RS)- and reading comprehension (RC) by presenting a text and a questionnaire, and evaluating the accuracy and time of the answers \cite{de2017longitudinal}. 
Finally, subjects were evaluated within the phonological deficit framework \cite{babiloni2012cortical,de2017longitudinal,goswamiNeuralOscillationsPerspective2019}, measuring categories such as PA and verbal short-term memory (VSTM), by counting the number of suppressed syllables or phonemes in a variety of listening tests; and the speed and accuracy on a rapid automatized naming (RAN) of objects, figures and words presented. These three measures conform the triad of tests of the classical theory of central phonological deficit.

Time and number of hits and misses were recorded during the tests, and a composite measure representing the throughput in the test answers (accuracy/time) was computed as $th = \frac{hit}{(hit + miss)t}$. Later, the percentile of $th$ was computed for each subject in the dataset in each test output. Finally, the percentiles of tests belonging to the same category (e.g., PA) were aggregated for each subject separately, to obtain an average measure of each category.

The researchers selected an EEG cohort composed of 48 subjects of age 7: 15 of which had clear reading deficit and 33 CTL. These groups had matching age and school-level socio-economic index (SEI) measured in a scale 1-10. The neuropsychological results of this cohort are given in Table \ref{tab:group_characteristics}, along with other demographic details.

\begin{table*}[h!]
\centering
\caption{Demographics and test results of the EEG cohort. Results of each test are
provided in terms of percentile within the whole 1400 LEEDUCA cohort and the standard
deviation per group is provided in brackets.}
\begin{tabular}{@{}lcc@{}}
\toprule
\textbf{Group}      & \textbf{CTL}      & \textbf{DLX}      \\ \hline
$N$                 & 33                      & 15                      \\ 
Age (months)        & 94.1 [3.3]              & 95.6 [2.9]              \\ 
SEI                 & 5.091 [2.53]            & 4.533 [2.07]            \\ 
\midrule
VR                  & 0.648 [0.20]            & 0.260 [0.20]            \\ 
PR                  & 0.602 [0.19]            & 0.233 [0.13]            \\ 
RS                  & 0.656 [0.18]            & 0.237 [0.18]            \\ 
RC                  & 0.645 [0.17]            & 0.375 [0.24]            \\ 
\midrule
PA                  & 0.616 [0.17]            & 0.275 [0.19]            \\ 
RAN                 & 0.517 [0.22]            & 0.378 [0.15]            \\ 
VSTM                & 0.468 [0.19]            & 0.389 [0.26]            \\ 
\botrule
\end{tabular}
\label{tab:group_characteristics}
\end{table*}

These students were presented with a white noise stimulus, AM at different frequencies: 4.8 Hz, 16 Hz and 40 Hz, corresponding to the Spanish mean frequencies of three speech units: syllable, intrasyllabic and phoneme \cite{Lizarazu2015}. The AM stimuli were presented in sequential ascending and descending order (4.8 - 16 - 40 - 40 - 40 - 16 - 4.8 Hz) for 2.5 minutes each, for a total acquisition time of 15 minutes. Acquisition of EEG signals was performed using a 10-20 configuration optimized for auditory processing. Vertical (V-) and horizontal (H-) electrooculogram (EOG) signals were also recorded for ocular correction. EEG data were preprocessed with EEGLAB, after artifact detection and removal, and divided into 5-second segments with no artifacts in any channel.

\subsection{Cross-Frequency Coupling}

CFC techniques have played a major role in EEG research in recent years \cite{Lizarazu2019phase}. With a solid theoretical basis, cross-frequency PAC has been linked to brain area communication involved in perception, cognition, and action \cite{sauseng2008cross,canolty2010functional}.

CFC measures the degree of coupling between two narrow-band signals. PAC relies on the computation of instantaneous phase ($\phi(t)$) and amplitude envelope \(a(t)\) signals at different frequencies. In our case, \(\phi_{1}(t)\)  is extracted from the synchronized stimulus, and \(a_{2}(t)\) is the amplitude from the EEG signal at different electrodes, after applying a FIR bandpass filter centered at \(\omega_i\). These signals are derived from the Hilbert transform (HT). The HT of any given signal \(x(t)\) is defined as: 

\begin{equation}
    \mathcal{H}[x(t)] = \frac{1}{\pi}  \int_{-\infty}^{+\infty} \frac{x(t)}{t-\tau} d\tau 
\end{equation}

\noindent Instantaneous phase and amplitude is then computed as the module and phase of a composite signal known as the analytic signal \(z(t)\), that is computed as: 

\begin{equation}
    z_{i}(t)=x_{i}(t)+j \mathcal{H} [x_{i}(t)] = a(t) e^{j\phi(t)}
\end{equation}

\noindent \(a(t)\) is then estimated as the module of \(z_{i}(t)\):

\begin{equation}
    a(t)=\sqrt{re(z_{i}(t))^2+im(z_{i}(t))^2}
\end{equation} 

\noindent whereas the instantaneous, unwrapped phase \(\phi(t)\) is estimated as: 

\begin{equation}
\label{ec:unwrapped_phase}
    \phi(t)=tan^{-1} \frac{im(z_{i}(t))}{re(z_{i}(t))}
\end{equation}

\noindent With these signals at hand, we can compute the different PAC measures reviewed at Section~\ref{sec:introduction}. The PLV \cite{mormann2005phase} is obtained from the phase of the lower-frequency signal \(\phi_1(t)\) and the phase of the higher-frequency envelope signal \(\phi_{a_2}(t)\) as: 

\begin{equation}
    PLV = \frac{1}{N}\sum_{t=0}^{N-1} \left|e^{-j\phi_{a_2}(t)} e^{-j\phi_1(t)}\right|
\end{equation} 

\noindent where \(N\) is the length of the signals for \(t\in 1 \dots N\).

For the HR \cite{lakatos2005oscillatory} and the KL-MI \cite{tort2008dynamic,tort2010measuring}, a vector quantization is applied on \(a_2(t)\) with respect to the phase signal \(\phi_1(t)\). To do so, the range of phases \([-\pi, \pi]\) is divided onto \(J\) quantiles, and therefore, a counting vector \(p_j\) is defined for the \(j^{th}\) quantile of phases in the instantaneous phase vector \(\phi_1(t)\): 

\begin{equation}\label{eq:vquant}
    p_j = \frac{1}{M}\sum_{t \in T_j} a_2(t) \quad \text{and} \quad T_j = \{t | \phi_1(t) \in j^{th} \text{quant.}\}
\end{equation} 

\noindent where \(M\) is the number of elements in the set \(T_j\).

In one of the first approaches to compute KL-MI \cite{tort2008dynamic}, we define the quantized vector \(\mathbf{p} = [p_0, \dots p_{J-1} ]\). Afterwards, \(\mathbf{p}\) is normalized so that it sums up to 1, so that its elements \(P_j\) can be interpreted as the probability of the \(j^{th}\) phase bin. KL-MI is computed as the Kullback-Leibler divergence of \(P\) with respect to the uniform distribution: 

\begin{equation}
    \textit{KL-MI} = 1 + \frac{1}{ \log(J)}\sum_{j=0}^{J-1} P_j \log(P_j)
\end{equation}

A simpler approach was the HR \cite{lakatos2005oscillatory}. Using the quantized vector \(\mathbf{p}\), the HR is
defined as: 

\begin{equation}
    HR = \frac{\max(\mathbf{p})-\min(\mathbf{p})}{\max(\mathbf{p})}
    \label{eq:hr}
\end{equation}

gcPAC \cite{ince2017statistical} is a continuous approach to compute Mutual Information (MI) on EEG signals. In \cite{ince2017statistical} it was proven that the MI between two continuous variables \(X\) and \(Y\) is equal to the negative entropy of their statistical copula. The method assumes the marginals to be standard Gaussian variables, and expresses the multivariate Cumulative Density Function (CDF) in terms of the marginal CDFs for each variable, and a copula function, according to Sklar's theorem. It is proven that the entropy of the joint distribution over \(X\) and \(Y\) equals the marginal entropy plus the entropy of the copula function. Therefore, the copula can be obtained as: 

\begin{equation}\label{gcPac}
    H(C) = H(X,Y)-H(X)-H(Y)
\end{equation}

The last method tested in this work is usually preferred for high-quality segments, the MVL \cite{canolty2010functional}. It is defined as the module of a signal whose amplitude is the envelope of the high-frequency signal \(a_2(t)\) and its phase is the phase of the low-frequency modulating signal \(\phi_1(t)\):

\begin{equation}\label{eq:mvl}
    MVL = \frac{1}{N}\sum_{t=0}^{N-1} \left| a_2(t) e^{j \phi_1(t)}\right|
\end{equation}

Once a PAC value has been computed for two vectors, it might be possible that high PAC was produced by chance. To rule out this possibility we use permutation testing to generate Z-scores, which give us an idea of the statistical significance of the observed PAC measure \(PAC_o\) as in \cite{tort2010measuring} and \cite{bahramisharif2013propagating}. In order to do so, we generate surrogate versions of the amplitude signal by barrel-switching the amplitude signal \(a_s(t)\) at different levels. Surrogate PAC measures are computed for each \(a_s(t)\), generating a surrogate PAC distribution from which Z-scores for the observed PAC (\(PAC_o\)) can be computed as:

\begin{equation}
    Z = \frac{PAC_o-\mu_{PAC_s}}{\sigma_{PAC_s}}
\end{equation} 

\noindent with \(\mu_{PAC_s}\) and \(\sigma_{PAC_s}\) the mean and standard deviation of the surrogate PAC distribution. In this work, 200 surrogates are generated for computing the PAC \(Z\)-score between the phase of the stimuli and the amplitude of EEG signals at different frequencies. This \(Z\)-score will be considered
henceforth as the PAC measure. PAC measures and permutation tests have been computed using the TensorPAC python package \cite{combrisson_tensorpac_2020}.

\subsection{Feature Aggregation and Clasification}

We estimated a single PAC \(Z\)-score measuring the interaction between a synchronized stimulus \(\phi(t)\) and a narrow-band filtered EEG signal at every electrode, \(a(t)\). \(a(t)\) was filtered at 49 frequency points ranging from \(\omega_\phi\), the frequency of the stimulus, to 200 Hz in logarithmic scale. This makes an array of \(31\times49\) PAC values per EEG segment, one for every combination of EEG channel and center frequency of the band pass applied to obtain \(a(t)\).

\subsubsection{Band PAC aggregation}

Spectral features such as the PAC are commonly aggregated in frequency bands \cite{LEHONGRE20111080,Lizarazu2015,molinaro2016out,Flanagan2018,DiLiberto2018}. In this work, we define the intervals of these bands as Delta (from 0.5 to 2 Hz), Theta (from 2 to 4 Hz), Alpha (from 4 to 8 Hz), Beta (from 8 to 25 Hz), Gamma (from 25 to 100 Hz) and Epsilon (from 100 to 200 Hz), also traditionally known as High-Gamma frequencies \cite{Lizarazu2015,molinaro2016out}.

\begin{figure}[h]
    \centering
    \includegraphics[width=0.8\columnwidth]{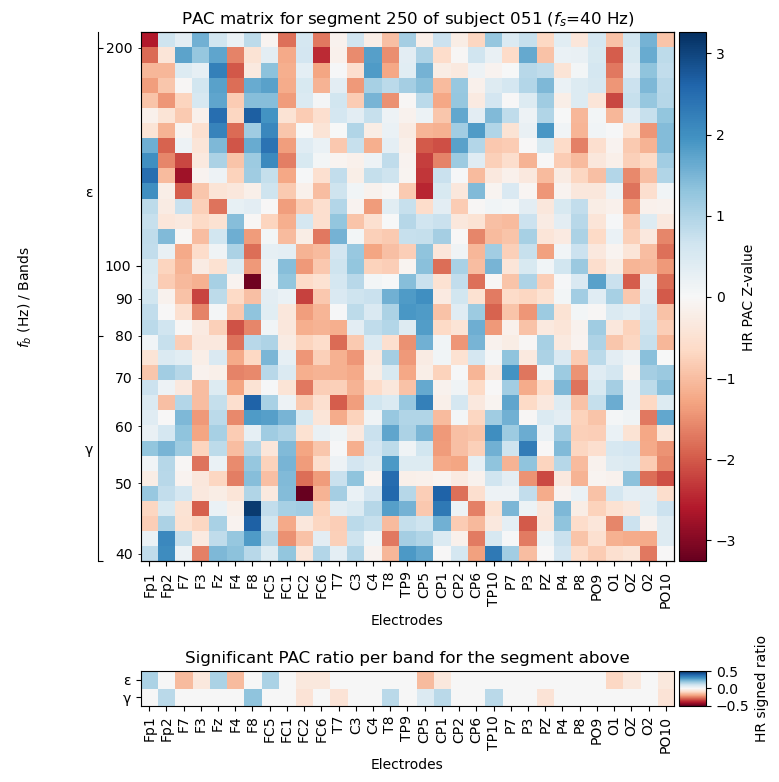}
    \caption{Illustration of the `ratio' band PAC aggregation method. The upper panel shows a heatmap of the original PAC data across frequencies and electrodes. The lower panel displays the significant PAC ratios aggregated by frequency band.}
    \label{fig:band_pac_aggr}
\end{figure}

The aggregation is performed in the following way. For each frequency band, a set of PAC values (in Z-score) is built \(PAC_{band} = \lbrace PAC_{f_i}, \dots \rbrace\) for all \(f_i \in (f_{ini}, f_{end})\), where \(f_{ini}\) and \(f_{end}\) are the frequency extrema of the bands (Figure \ref{fig:band_pac_aggr} upper panel). The final PAC for band \(B\) is the number of PAC values in the set \(PAC_{band}\) above a threshold \(Z_{th}=1.65\), which corresponds to the Z-score for that \(f\) being higher than 95\% of all values in the permutation set. Afterwards, this number can be normalized using the total number of elements in \(PAC_{band}\) (Figure \ref{fig:band_pac_aggr} lower panel). This is applied to each 5-second segment in the dataset. Afterwards, the \(PAC_{band}\) for every subject is obtained as the average of the PAC computed for every 5-second segment.

%

\subsubsection{Study of diagnostic accuracy using SVC}

To assess the diagnostic accuracy of the composite PAC-band features we used a linear SVC \cite{vapnikStatisticalLearningTheory1998}. SVCs are widespread in the machine learning literature \cite{martinez-murciaStudyingManifoldStructure2020}, particularly in their linear -basic- and radial basis kernel function (RBF), offering a remarkable trade-off between classification accuracy and generalization ability. In their basic, linear form, it is possible to observe the classification weights, providing a deeper interpretation of the relevance of the input features, even in probabilistic terms. Two distance metrics can be used in the computation of the SVC loss, the L1 and L2 norms. Whereas the latter usually leads to higher performance, the L1 norm tends to create more sparse matrices, focusing on a smaller number of features. Both alternatives will be tested in these experiments, using the python implementation of LIBSVM \cite{changLIBSVMLibrarySupport2011}.

Regarding SVC coefficients interpretation, the SVC linear discrimination model learns a model $\mathbf{y} = \mathbf{w}\mathbf{x}$, and therefore a coefficient vector $\mathbf{w}$ for the input data. In this work, the input data are the aggregated PAC Z-measures $B$ for each electrode and band, as commented before. A coefficient of the SVC is positive when the contribution of $B$ tilts the balance of the classifier towards predicting DLX subject, and negative when that contribution is towards CTL.

\subsubsection{Evaluation}

Our hypothesis states that control and dyslexic subjects differ in the way they process auditory stimuli, in a process commonly known as entrainment. This is in consonance with the phonological hypothesis for the origins of dyslexia and the TSF described in \cite{goswamiSpeechRhythmLanguage2019,martinezmurciaEEGConnectivityIJNS2020}. As a consequence, there may exist differences in PAC coupling between the stimuli presented to and the EEG signals acquired from the experimental subjects.

The classification is performed as follows. First, the features are standardized and divided into train and test sets. Then, a SVC classifier is trained. Repeated stratified 5-fold cross-validation is used to estimate the performance of the methodology. This implies that the dataset is randomly split in 5 sets, each of which has roughly the same proportion of classes (stratified). Four of them are aggregated to conform a training set, and the test set uses the remaining one. This procedure is repeated 50 times, in order to test different training and test sets.

\subsection{Estimation of the PAC Influence over Reading and Phonological Performance}

In order to explore the potential ability of PAC characteristics to accurately predict DLX in children, we explored the relationship between PAC measures and the categories of phonological and reading performance in which children were assessed. In addition, it improves the explainability of the proposed method.

The approach is based on Multi-task Lasso combined with permutation testing to compute the empirical statistical significance of the resulting coefficients. The Multi-task Lasso is an extension of Lasso regression developed specifically to address multiple interrelated tasks concurrently by taking advantage of shared patterns or structure between them. Specifically, it is a linear model that estimates sparse coefficients for multiple regression problems (task) jointly constrained by the requirement that the selected features have to be the same for all tasks. For this, it uses a mixed L1/L2 norm regularization that is shared across all task. In this way, it lead to a more stable feature selection compared to applying Lasso separately to each task. The objective function to minimize in Multi-task Lasso is:

\begin{equation}
    \min_{W} { \frac{1}{2n_{\text{samples}}} ||X W - Y||_{\text{Fro}} ^ 2 + \alpha ||W||_{21}}
\end{equation} 

\noindent where $X$ is the input data and $Y$ is the 2D matrix of target variables, in our case the aggregate PAC measures and the measures of reading and phonological performance, respectively. $||\cdot||_{\text{Fro}}$ indicates the Frobenius norm computed as
\begin{equation}
    ||A||_{\text{Fro}} = \sqrt{\sum_{ij} a_{ij}^2}
\end{equation}

\noindent $W$ is the coefficient matrix and $\alpha$ is the regularization parameter of the mixed L1/L2  norm for regularization denoted by $||\cdot||_{21}$ and defined as

\begin{equation}
    ||W||_{2 1} = \sum_i \sqrt{\sum_j W_{ij}^2}.
\end{equation}

\noindent in which $W_{ij}$ represents the regression coefficient for the $i^{th}$ feature in task $j$. The fitting of the model is based on a procedure known as coordinate descent \cite{hastie2015statistical} and we have employed the implementation of the Multi-task Lasso in the Python library scikit-learn \cite{pedregosaScikitlearnMachineLearning2011}. 

A permutation test has been computed by permuting the target reading and phonological performance variables. Null distribution is estimated by 99999 permutations of the dataset reading and phonological performance variables in a 5-fold cross validation scheme. Then, the p-values are obtained using a two-sided approach by calculating the fraction of permutations in which the Multi-task Lasso coefficients of the null distribution are greater or less than the observed Multi-task Lasso coefficients. The two sided alternative is selected to consider the positive and negative influence represented by the computed coefficients for each electrode. To account for the large number of simultaneous comparisons arising from the analysis across 31 EEG channels and 7 behavioural indices (217 tests per stimulus--band combination), the Benjamini--Hochberg False Discovery Rate (FDR-BH) procedure \cite{benjaminiControllingFalseDiscovery1995} was applied to the permutation-derived $p$-values. This method controls the expected proportion of false positives among the rejected hypotheses, offering greater statistical power than family-wise error rate corrections while maintaining principled control over spurious associations.

\backmatter

\bmhead{Acknowledgements}
This research is part of the {CNS2024-154296} action, funded by MICIU/AEI/10.13039/501100011033; the projects PID2022-137461NBC32, PID2022-137629OA-I00 and PID2022-137451OB-I00, funded by MICIU/AEI/10.13039/501100011033 and by ERDF/ EU, and TIC251-G-FEDER project, funded by ERDF/EU. This work has also been supported by  Leeduca research group (University of Málaga) along with data acquisition. Work by F.J.M.M. is part of the grant {RYC2021-030875-I} funded by MICIU/ AEI/10.13039/501100011033 and by European Union NextGenerationEU/PRTR.

\section*{Declarations}

Some journals require declarations to be submitted in a standardised format. Please check the Instructions for Authors of the journal to which you are submitting to see if you need to complete this section. If yes, your manuscript must contain the following sections under the heading `Declarations':

\begin{itemize}
\item Conflict of interest/Competing interests 
The authors declare no competing interests.

\item Ethics approval and consent to participate
This study was conducted in accordance with the Declaration of Helsinki and approved by the Medical Ethics Committee of the University of Málaga (approval reference: PND016/2020, dated 05/02/2020). The study was additionally authorised by the Education Office of the Junta de Andalucía, which granted permission to evaluate students at participating public schools. Prior to enrolment, written informed consent was obtained from the parent or legal guardian of each participating child. All procedures involving human participants were approved and conducted under the oversight of the LEEDUCA research group at the University of Málaga.
\item Author contribution
F.J.M.-M. and A.O. contributed equally to the conceptualization and design of the study. F.J.M.-M. and M.A.F. designed and implemented the EEG signal processing pipeline and PAC analysis software. F.J.M.-M. and N.J.G.-M. performed the formal analysis and interpretation of results. A.O. was responsible for data acquisition and coordination of the LEEDUCA cohort. F.J.M.-M., N.J.G.-M. and A.O. wrote and revised the manuscript. A.O. and F.J.M.-M. provided project supervision. F.J.M.-M. and A.O. were responsible for funding acquisition.

\end{itemize}

\bibliography{bibliography}

\end{document}